\documentclass[american,aps,prx,superscriptaddress,reprint,longbibliography]{revtex4-1}
\usepackage{graphicx,color}  
\graphicspath{{figures/}}
\usepackage{wrapfig}
\usepackage{dcolumn}   
\usepackage{bm}        
\usepackage{amssymb}   
\usepackage{amsmath} 
\usepackage{subfig}
\newcommand{\ket}[1]{\vert #1 \rangle}

\newcommand{\outpr}[2]{\vert {#1} \rangle \langle {#2} \vert}

\DeclareMathOperator{\sinc}{sinc}
\DeclareMathOperator{\expc}{exp}
\usepackage[unicode=true,
 bookmarks=true,bookmarksnumbered=false,bookmarksopen=false,
 breaklinks=false,pdfborder={0 0 0},pdfborderstyle={},backref=false,colorlinks=true]
 {hyperref}
\hypersetup{
 pdfborderstyle={},allcolors=blue}

\begin{document}


\title{Fast Simulation of Magnetic Field Gradients for Optimization of Pulse Sequences}

\author{John P. S. Peterson}

\thanks{e-mail:  johnpetersonps@hotmail.com}

\affiliation{Institute for Quantum Computing and Department of Physics and Astronomy,
University of Waterloo, Waterloo N2L 3G1, Ontario, Canada}
                        
\author{Hemant Katiyar}

\affiliation{Institute for Quantum Computing and Department of Physics and Astronomy,
University of Waterloo, Waterloo N2L 3G1, Ontario, Canada}

\author{Raymond Laflamme}

\affiliation{Institute for Quantum Computing and Department of Physics and Astronomy,
University of Waterloo, Waterloo N2L 3G1, Ontario, Canada}

\affiliation{Perimeter Institute for Theoretical Physics, 31 Caroline Street North, Waterloo, Ontario, N2L 2Y5, Canada}

\affiliation{Canadian Institute for Advanced Research, Toronto, Ontario M5G 1Z8, Canada}
\date{\today}

\begin{abstract}
        We study how to simulate, efficiently, pulse field gradients (PFG) used in nuclear magnetic resonance (NMR). 
        An efficient simulation requires discretization in time and space. We study both discretizations and provide a 
        guideline to choose best discretization values depending on the precision required experimentally.
        We provide a theoretical study and simulation showing the minimum number of divisions we need in space for simulating, 
        with high precision, a sequence composed of several unitary evolution and PFG. We show that the fast simulation of 
        PFG allow us to optimize sequences composed of PFG, radio-frequency pulses and free evolution, to implement non-unitary 
        evolution (quantum channels). As an evidence of the success of our work, we performed two types of experiments. First,
        we implement two quantum channels and compare the results with their theoretical predictions. In the second experiment, 
        we used the fast simulation of PFG to optimize and implement a sequence to prepare pseudo pure state with better signal 
        to noise ratio than any known procedure till now. 
\end{abstract}

\maketitle

\section{Introduction}

In the 70s Lauterbur, Mansfield and Maudsley showed that magnetic field gradients, together with the magnetic resonance technique, 
could be used to produce images \cite{Lauterbur,Mansfield1,Mansfield2,Mansfield3}. 
Magnetic resonance imaging (MRI) became a very useful technique in medical science ever since.
In high-resolution nuclear magnetic resonance (NMR) spectroscopy, one of the first notable applications of the magnetic field gradient 
was presented by Stejskal and  Tanner. They demonstrated how gradients can be used to determine the diffusion coefficient of liquids \cite{Stejskal}. 
With time as the equipments used to produce the gradients started improving \cite{equip1,equip2,equip3,equip4,equip5,equip6}, 
new applications emerged \cite{seq1,seq2,seq3,seq4,seq5,seq6,seq7,seq8,seq9}.
Recently, gradients were used in quantum thermodynamics experiments to implement measurement protocols \cite{szi,ter1} 
and to prepare thermal states \cite{ter2,ter3}.
In NMR quantum computing, coherence pathway selection technique \cite{livrole}, which uses gradients, 
is routinely employed to transform the thermal state of a group of nuclear spins into a pseudo pure state \cite{livro}.
Magnetic field gradients were also used to simulate noise \cite{oti,dco}. Given numerous applications, 
a good theoretical description and methods to simulate the dynamics of the nuclear spins under the influence of magnetic field gradients 
are becoming increasingly essential to design new experiments.

Nowadays, we already know few ways to simulate the dynamics of spins under the influence of magnetic field gradients, 
and some software and algorithms are already available \cite{rgrad0,rgrad1,rgrad2,rgrad3,rgrad4,rgrad5,rgrad6,rgrad7,rgrad8}. 
Among these methods, here we will be interested in the one presented by Allard \textit{et. al.} \cite{rgrad0}. In this method, the space 
is discretized into several divisions, and in each division we have a spin that due to the magnetic field gradient has its oscillation frequency slightly modified. 
Thus, when applying a magnetic field gradient, the final state of the system will be given by the average of the final states of the spins in each division. 
This method has the advantage of completely describing the final state of the system, but it can be slow when used to study the dynamics of molecules composed of many nuclear spins. However, as we will demonstrate here, we can accelerate this method using approximations and/or specific configurations that
allows us to perform the simulation with a small number of divisions. 

Here, we show that with an optimum value of number of divisions, it is possible to simulate quickly and with high precision sequences composed of several radio-frequency pulses, free evolutions and gradients. We also show that, together with optimization algorithms \cite{livrooti}, a fast simulation of the dynamic of spins under the influence of gradients can be used to optimize non-unitary evolutions, which are essential for implementing quantum channels or preparing specific states \cite{livronc}. We carried out experiments, implementing two quantum channels, to demonstrate that your simulations describe the dynamics of the system with good accuracy. Finally, we also use our results to optimize and prepare experimentally a pseudo pure state with better signal to noise ratio. During the optimization of pseudo pure state, we saw a trend pointing to a limit for the maximum improvement in the signal to noise ratio.

This paper is organized as follows: in section \ref{sec:Theory} we review NMR theoretical background. Then, we describe the gradient discretizations in time and space in section \ref{sec:simulation}. We study time discretization in section \ref{sec:TimeDisc} and space discretization in section \ref{sec:SpaceDisc} to provide
a guideline for choosing efficient discretization values.
Finally, in section \ref{sec:experiment}, we use the techniques developed to optimize sequence and test it experimentally.

\section{NMR theory}\label{sec:Theory}

Here, we are going to consider that our sample is an isotropic liquid, but our results can be generalized to other types of samples.
For the samples that we used in your experimental test, we can study the natural dynamics of the system using the following Hamiltonian:
\begin{equation}\label{eq:h0}
 \begin{split}
\ \mathcal{H}_{0} = \sum_{k}\frac{\hbar(\omega_{k}-\omega_{R})\sigma_{z_{k}}}{2} + \sum_{k \neq n}\frac{\pi\hbar J_{kn}\sigma_{z_{k}}\sigma_{z_{n}}}{4},
 \end{split}
 \end{equation} 
where $\omega_{k}$ and $\sigma_{\beta_{k}}$ are, respectively, the angular oscillation frequency and the Pauli matrix $\beta$ of the $k$-th nuclear spin, $\hbar$ is the Planck constant divided by $2\pi$, $\omega_{R}$ is the angular frequency of the rotating frame \cite{livrole} and $J_{kn}$ is the scalar coupling constant of the spins $k$ and $n$. 

If we apply a magnetic field gradient, whose magnitude increase linearly along the $z$ direction, the Hamiltonian of the nuclear spins of a molecule in the position $z$ at time $t$ will be given by:
\begin{equation}\label{eq:hg}
 \begin{split}
\ \mathcal{H}_{\texttt{g}}(t,z) = \sum_{k}\frac{ \hbar \gamma_{k}  \texttt{g}(t) z\sigma_{z_{k}}}{2},
 \end{split}
 \end{equation} 
where $\gamma_{k}$ is the gyromagnetic ratio of the $k$-th nuclear spin and $\texttt{g}(t)$ is the magnitude of the magnetic field gradient at time $t$. In this work, we are considering the case where the duration of the applied gradient is fast enough so that we can have a good description of the system dynamics, without including the diffusion or relaxation processes. 
 
The nuclear spins state are controlled by radio-frequency pulses applied in the $xy$ plane with an angular frequency $\omega_{R}$. The Hamiltonian that describes the interactions of the spins with a pulse in the rotation frame will be given by:
\begin{equation}\label{eq:hc}
 \begin{split}
\ \mathcal{H}_{c}(t) = \hbar \Omega(t) \sum_{k=1}^{s}\frac{\cos[ \phi(t) ] \sigma_{x_{k}} + \sin[ \phi(t) ] \sigma_{y_{k}}}{2},
 \end{split}
 \end{equation} 
where $\Omega(t)$ and $\phi(t)$ are the modulations of the
pulse amplitude and phase respectively.

Considering the interactions described above, the total Hamiltonian of the nuclear spins from a molecule in the position $z$ at time $t$ is given by:
\begin{equation}\label{eq:ht}
 \begin{split}
\ \mathcal{H}_{T}(t,z) = \mathcal{H}_{0} + \mathcal{H}_{\texttt{g}}(t,z) + \mathcal{H}_{c}(t),
 \end{split}
 \end{equation} 
and the evolution of this system under the action of $\mathcal{H}_{T}(t,z)$ will produce the following unitary:
\begin{equation}\label{eq:uht}
 \begin{split}
\ U_{\mathcal{H}_{T}}(z) = \mathcal{T} \left[ \exp\left(-\frac{i}{\hbar}  \int \mathcal{H}_{T}(t,z) dt \right) \right],
 \end{split}
 \end{equation} 
where $\mathcal{T}$ represents the Dyson time-ordering operator.

\section{Gradient simulation} \label{sec:simulation}
 
In our simulation, we consider that the molecules are uniformly distributed along the $z$ axis, and they are diluted so that we can disregard intra-molecular interactions. A molecule in the position $z$ will evolve under the Hamiltonian $\mathcal{H}_{T}(t,z)$, eq.~\eqref{eq:ht}, and after time $t$, the state of the nuclear spins of this molecule will be given by:
\begin{equation}\label{eq:estado}
 \begin{split}
\ \rho(t,z) =  U_{\mathcal{H}_{T}}(z)\rho (0,z)U^{\dagger}_{\mathcal{H}_{T}}(z),
 \end{split}
 \end{equation} 
where $\rho (0,z)$ represents the initial state of these spins. The state of the whole sample will be given by:
\begin{equation}\label{eq:estadot1}
 \begin{split}
\ \rho_{S}(t) =  \dfrac{\int \rho(t,z) dz}{\int dz}.
 \end{split}
 \end{equation} 
We need to perform two types of discretizations to simulate the dynamics of the system. One in the time, to calculate $U_{\mathcal{H}_{T}}(z)$, and another in the space, to obtain the value of $\rho_{S}(t)$. It is worth mentioning that our goal is not to present an accurate method, but one where an approximate simulation of the system dynamics can be obtained quickly.
  
\begin{figure}[h]%
    \includegraphics[width=8.5cm]{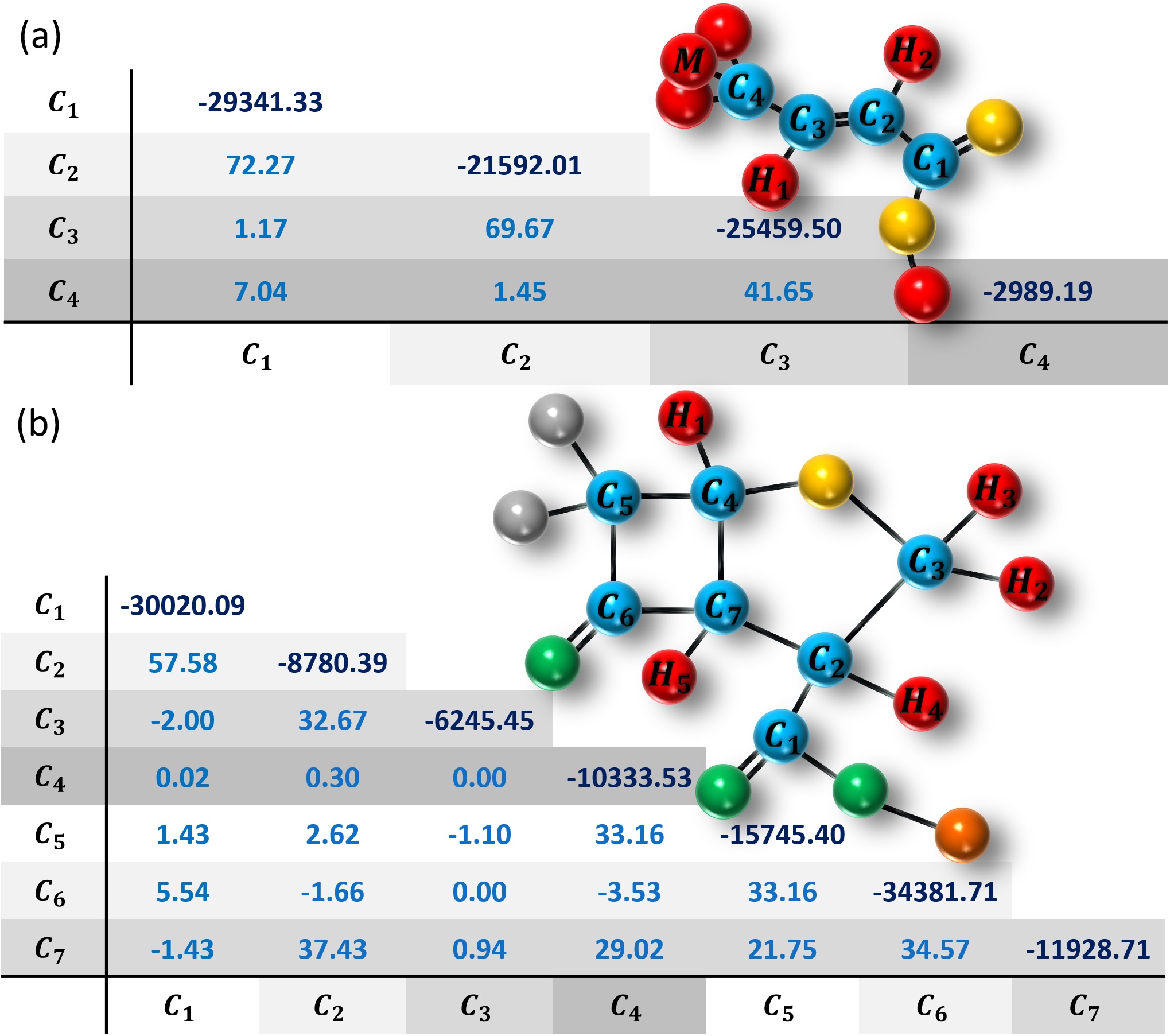}
    \centering
    \caption{ Sample information for (a) $^{13}\textrm{C}$-labeled transcrotonic acid molecule (4 qubits system) and (b) per-$^{13}\textrm{C}$-labeled (1S,4S,5S)-7,7-dichloro-6-oxo-2-thiabicyclo[3.2.0]heptane-4-carboxylic acid molecule (7 qubits system) - The off-diagonal terms in the table are the $J$ coupling constants of the $^{13}\textrm{C}$ nuclear spins of the molecules. Meanwhile, on the diagonal we have the values of the chemical shifts of each nuclear spin. The values in the table are in Hz.}%
    \label{fig:molecula4q}%
\end{figure}

To provide a realistic estimate of these approximations, we use two types of molecules for our simulation:
the $^{13}\textrm{C}$-labeled transcrotonic acid and the per-$^{13}\textrm{C}$-labeled (1S,4S,5S)-7,7-dichloro-6-oxo-2-thiabicyclo[3.2.0]heptane-4-carboxylic acid. The $^{13}\textrm{C}$ nuclear spins have spin-$1/2$
and thus the molecules can be used to represent physically a set of 4 and 7 qubits, respectively.
The values of the resonance frequencies and the scalar coupling constants of the $^{13}\textrm{C}$ nuclear spins of the two molecules are shown in fig. \ref{fig:molecula4q}(a-b).

We use fidelity \cite{livronc} as a measure of the distance between the final states obtained with and without the use of approximations in the simulation. 
When performing simulations, we considered that the molecules are uniformly distributed in the $z$-direction
and when the field gradient is applied, each molecule will have a slightly different resonance frequency given by their physical location.
Due to hardware restrictions, some NMR equipments requires delays of a few $\mu$s before and after the application of a field gradient, in our simulations this delay is $200 \mu$s.

\subsection{Time discretization}\label{sec:TimeDisc}

We discretize the total time of evolution $\tau$ into $m$ intervals of duration $\delta t$. The value of $\delta t$ must be small enough to allow us to consider that $\mathcal{H}_{T}(\delta t,z)$ is approximately constant at each of the $m$ time intervals. Then, the value of $U_{\mathcal{H}_{T}}(z)$ can be calculated by:
\begin{equation}\label{eq:uhtpro}
 \begin{split}
\ U_{\mathcal{H}_{T}}(z) = U_{m}(z)U_{m-1}(z)U_{m-2}(z) \cdots U_{2}(z)U_{1}(z), 
 \end{split}
 \end{equation} 
with
\begin{equation}\label{eq:uhtprok}
 \begin{split}
\ U_{k}(z) =  \exp\left\lbrace-\frac{i}{\hbar} \mathcal{H}_{T}(k \delta t,z)  \delta t \right\rbrace .
 \end{split}
 \end{equation}
 
One of the most time-consuming computational operations in this simulation is the computation of the exponential of the matrix present in eq. \eqref{eq:uhtprok}. Therefore, we must adopt strategies to calculate the value of $U_{k}(z)$ efficiently.
The strategy used will depend on whether radio-frequency pulses are applied during the implementation of the magnetic field gradient.

If we consider that pulses are not applied together with the gradient, the Hamiltonian of the system during the gradient will always be diagonal in the $\sigma_{z}$ basis. Thus, we do not need to calculate the exponential of matrices, because $U_{k}(z)$ can be determined by calculating the exponential of the diagonal elements of $-i \mathcal{H}_{T}(k \delta t,z)  \delta t /\hbar$. 
Since $U_k$ and $U_{\mathcal{H}_{T}}(z)$ are diagonal, we can determine the $j^{th}$ diagonal  element of $U_{\mathcal{H}_{T}}(z)$ by multiplying all the $j^{th}$ diagonal elements of the $m$ matrices $U_k$. By doing this, we do not need to perform the matrix multiplications from eq. \eqref{eq:uhtpro}.

In the special case where the amplitude of the gradient does not depend on time and we do not apply pulses during the gradient,
the total Hamiltonian, $\mathcal{H}_{T}$, will be independent of time too. If the gradient is applied for a time $\tau$, the evolution is given by a simplified equation:
\begin{equation}\label{eq:uhtsemt}
 \begin{split}
\ U_{\mathcal{H}_{T}}(z) = \exp\left\lbrace-i \mathcal{H}_{T}(z)  \tau /\hbar \right\rbrace.
\end{split}
 \end{equation}
Although this case has several restrictions, it is widely used in experiments of quantum computing, quantum information and thermodynamics.

In the case where pulses are applied together with the gradient, the system's Hamiltonian is not always diagonal. However, for systems composed only of spins $1/2$ (qubits), we can avoid the matrix exponentiation if we use the approximation presented by Bhole and Jones \cite{apro} with a slight modification to include the magnetic field gradient. This approximation requires a small discretization in time, $\delta  t$, to calculate the value of $U_{k}(z)$ with a good precision. According to this approximation, for a system composed of $Q$ qubits, we can write
\begin{equation}\label{eq:uhtprokaproximado}
 \begin{split}
\ U_{k}(z) \approx W_{k}^{+}(z)H_{Q} e^{-i\Omega (k\delta t)\varsigma  \delta t} H_{Q}W_{k}^{-}(z),
 \end{split}
 \end{equation}
where $H_{Q}$ is the tensor product of $Q$ Hadamard gates and $W_{k}^{\pm}(z)=e^{-i [\mathcal{H}_{0} +\mathcal{H}_{g}(k \delta t,z) \pm 2\phi (k\delta t)\varsigma  /\delta t ] \delta t/2}$, with $\varsigma  = \sum_{l=1}^{Q}\sigma_{z_{l}}/2$. Since the matrix $\varsigma $ is diagonal, the value of $U_{k}(z)$ can be determined without a need of matrix exponentiation. 

\begin{figure}[h]%
    \includegraphics[width=8.5cm]{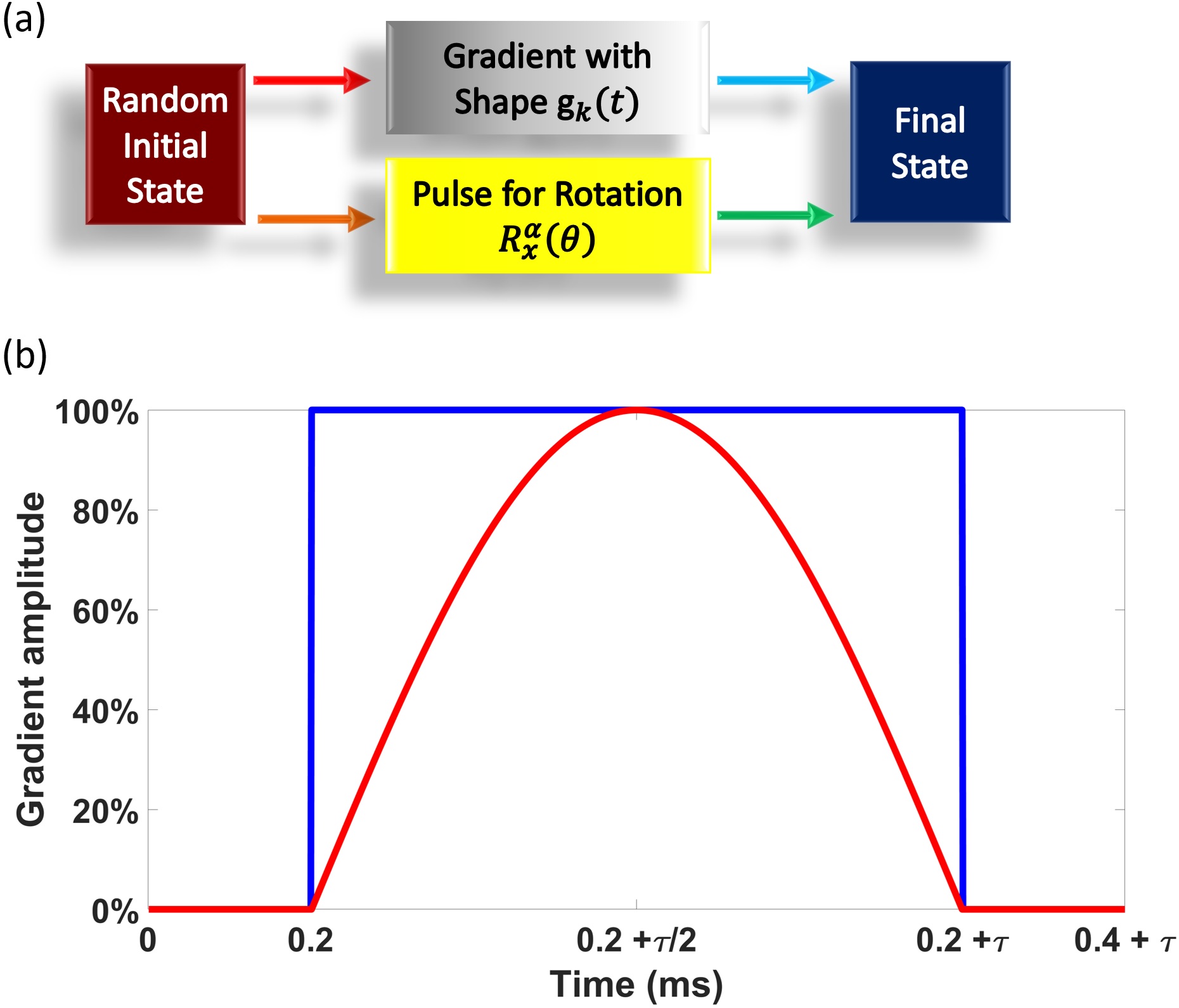}
    \centering
    \caption{ Sequence and shape of the amplitude of the magnetic field gradient - (a) sequence used to analyse the error due to the approximation used in  eq. \eqref{eq:uhtprokaproximado}. (b) Shape of the amplitude of the magnetic field gradient. The blue line is the graphical representation of the function $\texttt{g}_{1}(t)$, and the red one is the representation of the function $\texttt{g}_{2}(t)$, with $\texttt{g}_{2}(t) = sin[\pi t/(0.2+ \tau)]$ for $ 0.2 \leq t \leq 0.2+\tau $.}%
    \label{fig:seqgr}%
\end{figure}

In order to study the errors due to the approximation presented in eq. \eqref{eq:uhtprokaproximado}, 
we start with a random initial state, apply an unitary operator and a field gradient simultaneously, and calculate the a final state using the approximation from eq. \eqref{eq:uhtprokaproximado}.
We compare this final state with the final state when we
do not use the approximation, for different values of $\delta t$.
A graphical representation of this scheme is shown in fig. \ref{fig:seqgr}(a).
In our simulations, we start with 1024 different random initial states. Then, we apply in each of these states a field gradient, whose amplitude is modulate by one of the two shapes $\texttt{g}_{k}(t)$ shown in fig. \ref{fig:seqgr}(b), and one of the rotations: $R_{x}^{all}(\pi /2)$, 
$R_{x}^{odd}(\pi /2)$ and $R_{x}^{odd}(\pi)$, where $R_{x}^{\alpha}(\theta)$ is a rotation of an angle $\theta $ around the 
axis $x$, in the nuclear spins $\alpha$. Thus, resulting in six simulations for each random initial state.

\begin{figure}[h]%
    \includegraphics[width=8.5cm]{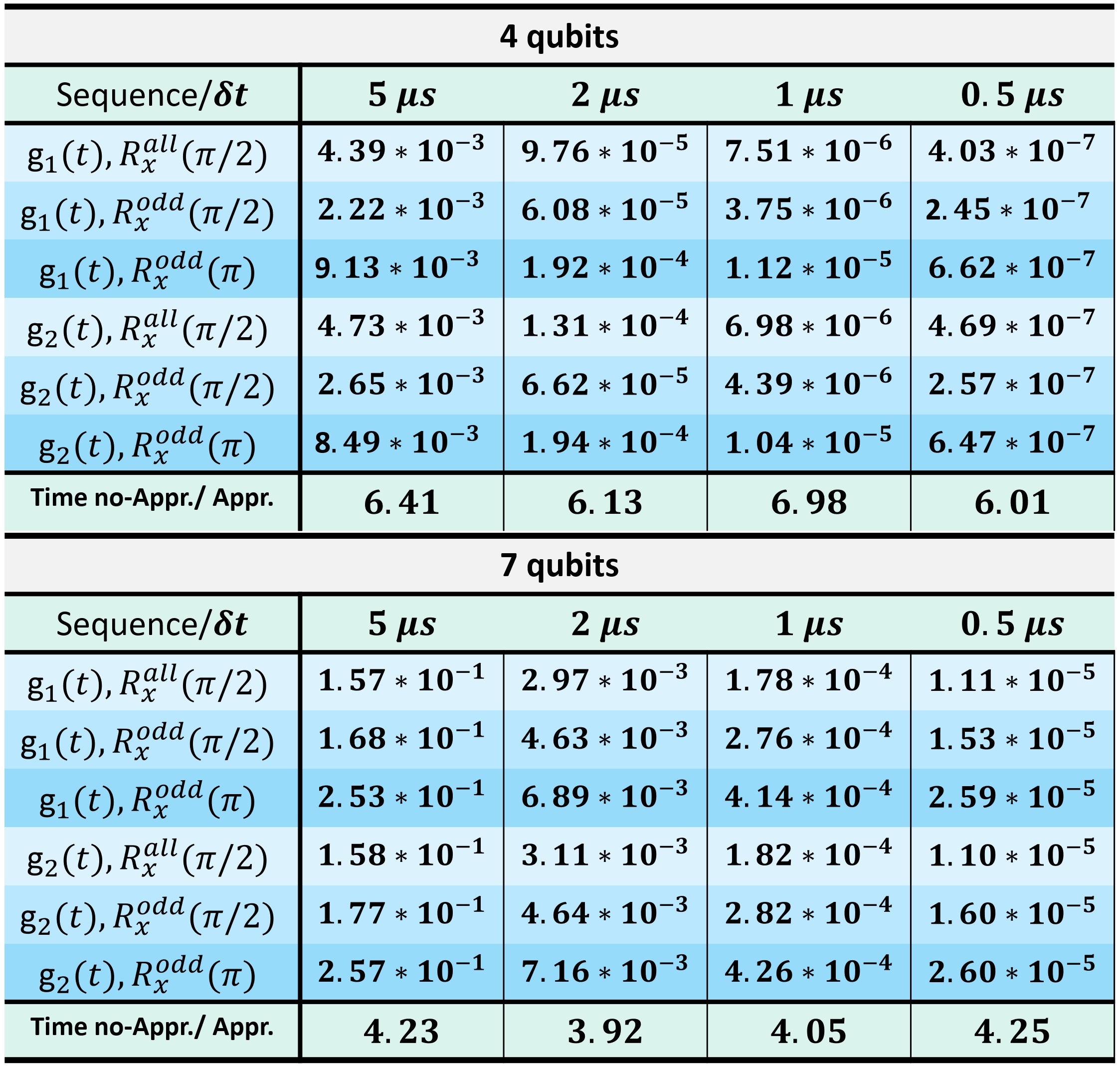}
    \centering
    \caption{ Error due to the approximation used in eq. \eqref{eq:uhtprokaproximado} - The error was estimated for the 4 and 7 qubits system, considering different pulses, shapes for the gradient and 1024 random initial states. The values of best infidelity ($1-worst(fidelity)$) are show for different discretization, gradient shape and rotations. }%
    \label{fig:tab1}%
\end{figure}

The pulses used to apply the rotations were optimized using the method developed by Peterson \textit{et. al.} \cite{johnp}. 
For simulating gradient with and without the approximation from eq. \eqref{eq:uhtprokaproximado}, we used an ensemble with $N = 10^4$ to minimize errors due to space discretization. The length of the sample, $L$, was considered to be $5$ cm. The pulse and the field gradient are applied simultaneously and have a duration, $\tau = 500$ $\mu$s. 
We performed simulations to estimate the error of four different values of time dicretization,  $ \delta t = \left \{5,2,1,0.5\right \}$ $\mu$s.
After these simulations, we compared the fidelity between the final states obtained with and without the approximation of eq. \eqref{eq:uhtprokaproximado}. 
In fig. \ref{fig:tab1}, we report the worst fidelity obtained among the 1024 initial states for the 4 and 7 qubits system as well as the ratio of simulation
time with and without eq. \eqref{eq:uhtprokaproximado} for different values of $ \delta t $.

As we can see in fig. \ref{fig:tab1}, the fidelity does not vary significantly when the shape of the gradient or the rotation are changed. 
However, we can have a big variation when $\delta t$ is changed. 
This gives us a way to choose the minimum value of $\delta t$ that satisfies a desired precision. For example, if our goal is to perform a simulation
with fidelity $0.99999$ for a $4$ qubit system, we use $\delta t=$ 1 $\mu$s in eq. \eqref{eq:uhtprokaproximado} and obtain the result faster than not using the approximation. 
For this case, the simulation (with $\delta t=$ 1 $\mu$s and using eq. \eqref{eq:uhtprokaproximado}) will be faster by a factor of 6.98, 3.49 or 1.396 if we compare with the time of the simulation without approximation and using $\delta t=$ 1 $\mu$s, $\delta t=$ 2 $\mu$s or $\delta t=$ 5 $\mu$s, respectively.
In fig. \ref{fig:tab1}, we can see that the precision also depends strongly on the system used. Thus, if the system used is different from the two considered here, the fig. \ref{fig:tab1} must be reconstructed for this new system. Once the values are characterized for a new system, they can be used for different experiments.

\subsection{Space discretization}\label{sec:SpaceDisc}

When we discretize the space, the integral in eq. \eqref{eq:estadot1} is replaced by a sum and the state of the ensemble divided into $N$ division with each division comprised of same number of molecules. Then, the state of the whole sample will be given by the following sum:
\begin{equation}\label{eq:estadot}
 \begin{split}
\ \rho_{S}(t) =  \dfrac{\sum_{k=1}^{N} \rho(t,k \delta z)}{N},
 \end{split}
 \end{equation} 
where $\delta z$ is the size of the discretization of the space. Generally, NMR samples are prepared in cylindrical tubes that are filled with liquids up to a height $L$, then we have $\delta z = L/N$.
 
Our goal is to estimate the smallest number of divisions, $N$, for us to be able to simulate quickly and with a high precision the dynamics of the system when we apply magnetic field gradients.
Here, we will consider that pulses are not applied simultaneously with the magnetic field gradient. 
This will facilitate our analysis and will allow us to avoid the errors due to the approximation presented in eq. \eqref{eq:uhtprokaproximado}.

\subsubsection{Density matrix}
A density matrix can be written as a summation of individual terms:

\begin{equation}\label{eq:basis}
\rho  = \sum_{v,w\in [1,2^{Q}]}a_{vw}\outpr{b_v}{b_w},
\end{equation}
where $b_v$ is the binary number ($v-1$) of length $Q$, with $v \in \{1,2,...,2^{Q}\}$. For example, if $Q=2$, then $b_1 = 00, b_2 = 01, b_3 = 10 $ and $b_4 = 11$.

\subsubsection{Order of coherence}
Coherence terms in a density matrix correspond to transition between different states, 
and the order of coherence is defined by how much is the change in the spin angular momentum quantum number, $m_l$. 
$\ket{0},\ket{1}$ are the eigenstates of $\sigma_z$ with eigenvalues $+1$,$-1$ and angular momentum quantum number  $m_l = +\frac{1}{2}$ and $-\frac{1}{2}$,
respectively. Then, $\outpr{00}{10}$ and $\outpr{01}{11}$ have coherence order, 
$\Delta m_l = -1$, $\outpr{11}{00}$ have coherence order, $\Delta m_l = 2$. For $Q$ qubits, coherence order can vary from -$Q$ to $Q$.

A term of the form $\outpr{b_v}{b_w}$ will have coherence order: 
\begin{equation}\label{eq:coh}
c_{vw}=\frac{1}{2}\sum_{k=1}^{Q} [(-1)^{b_{w}^{k}}-(-1)^{b_{v}^{k}}],
\end{equation}
where, $b_{v}^{k}$ is the $k^{th}$ element of binary number $b_v$. For example, if $b_v = 01$, we will have $b_{v}^{1}=0$ and $b_{v}^{2}=1$.

\subsubsection{Evolution during a gradient}
Since $\mathcal{H}_0$ and $\mathcal{H}_{\texttt{g}}(t,z)$ commute, we can write the evolution of the nuclear spin at position $z$ as:
\begin{eqnarray}
 U_{\mathcal{H}_{T}}(z) = U_0 \cdot U_{\texttt{g}}(z)  = U_{\texttt{g}}(z)\cdot U_0,
\end{eqnarray}
where $U_0$ and $U_{\texttt{g}}(z)$ are evolution under $\mathcal{H}_0$ and $\mathcal{H}_{\texttt{g}}(t,z)$ respectively.
Here, we consider that the gradient is time independent, $\texttt{g}_{k}(t)=\texttt{g}$, and the system is homonuclear, \textit{i.e.}, $\gamma_k = \gamma$. However, our analysis can be extend to time dependent gradient amplitude and heteronuclear systems.
The evolution of a term of the form $\outpr{b_v}{b_w}$ will result in:
\begin{eqnarray}\label{eq:evo}
        U_{\mathcal{H}_{T}}(z)\outpr{b_v}{b_w}&& U_{\mathcal{H}_{T}}(z)^{\dagger} \nonumber \\
        &&= A_{vw} U_{\texttt{g}}(z)\outpr{b_v}{b_w} U_{\texttt{g}}(z)^{\dagger},
\end{eqnarray}
where $A_{vw}$ is the constant produced by the application of $U_0$ in $\outpr{b_v}{b_w}$. The exact value of $A_{vw}$ can be easily calculated for small systems.
After the evolution under the gradient (see appendix \ref{ap:a1}), and using eq. (\ref{eq:coh}), the total evolution is given by:
\begin{eqnarray}
  U_{\mathcal{H}_{T}}(z)\outpr{b_v}{b_w}&& U_{\mathcal{H}_{T}}(z)^{\dagger} \nonumber \\
        &&= A_{vw}\expc\left(-i\gamma \texttt{g} z t c_{vw}\right) \outpr{b_v}{b_w}. \label{eq:grevo1}
\end{eqnarray}

Using eq. (\ref{eq:estado}), eq. (\ref{eq:basis}) and eq. (\ref{eq:grevo1}) the density matrix of the spins at position $z$ at time $t$,
\begin{equation}\label{eq:dde}
 \begin{split}
\ \rho(t,z) = \sum_{v,w\in [1,2^{Q}]}a_{vw} A_{vw}\expc\left(-i\gamma \texttt{g} z t c_{vw}\right) \outpr{b_v}{b_w}.
 \end{split}
 \end{equation}
 For infinitely many divisions of the sample of length $L$, hereby referred as the continuous case, we can use eq. (\ref{eq:estadot1}) to describe the state of the whole sample as (appendix \ref{ap:a1}):
\begin{widetext}
\begin{eqnarray}
        \rho_{S}(t)= \sum_{v,w\in [1,2^{Q}]} a_{vw} A_{vw}\Big(\sinc(\gamma \texttt{g} L t c_{vw}) - i \sinc(\gamma \texttt{g} L t c_{vw}/2)\sin(\gamma \texttt{g} L t c_{vw}/2) \Big)\outpr{b_v}{b_w}. \label{eq:continous2}
\end{eqnarray}
\end{widetext}
When dividing sample into a finite number of divisions, hereby referred as the discrete case, using eq. (\ref{eq:estadot}) we obtain (appendix \ref{ap:a1}),
\begin{widetext}
        \begin{eqnarray}
        \rho_{S}(t)= \sum_{v,w\in [1,2^{Q}]}a_{vw} A_{vw}\left(\frac{\cos(\frac{\gamma \texttt{g}L t c_{vw}}{2}\frac{N}{N-1})\sin(\frac{\gamma \texttt{g} Lt c_{vw}}{2})}{N\sin(\frac{\gamma \texttt{g} Lt c_{vw}}{2}\frac{1}{N-1})} + i
\frac{\sin(\frac{\gamma \texttt{g} Lt c_{vw}}{2}\frac{N}{N-1})\sin(\frac{\gamma \texttt{g}L t c_{vw}}{2})}{N\sin(\frac{\gamma \texttt{g}L t c_{vw}}{2}\frac{1}{N-1})}
                \right) \outpr{b_v}{b_w}. \label{eq:discrete}
\end{eqnarray}
\end{widetext}
For large $N$, $ \frac{N}{N-1} \approx 1 $ and 
$ \sin(\frac{\gamma \texttt{g}L t c_{vw}}{2}\frac{1}{N-1}) \approx  \frac{\gamma \texttt{g}L t c_{vw}}{2}\frac{1}{N-1} $. Hence,
\begin{widetext}
\begin{eqnarray}
\frac{\cos(\frac{\gamma \texttt{g}L t c_{vw}}{2}\frac{N}{N-1})\sin(\frac{\gamma \texttt{g} Lt c_{vw}}{2})}{N\sin(\frac{\gamma \texttt{g} Lt c_{vw}}{2}\frac{1}{N-1})} \approx
\frac{\cos(\frac{\gamma \texttt{g}L t c_{vw}}{2})\sin(\frac{\gamma \texttt{g} Lt c_{vw}}{2})}{\frac{\gamma \texttt{g} Lt c_{vw}}{2}} = 
\frac{\sin(\gamma \texttt{g} Lt c_{vw})}{\gamma \texttt{g} Lt c_{vw}} = \sinc(\gamma \texttt{g} Lt c_{vw}), \label{eq:real}
\end{eqnarray} 
and
\begin{eqnarray}
\frac{\sin(\frac{\gamma \texttt{g} Lt c_{vw}}{2}\frac{N}{N-1})\sin(\frac{\gamma \texttt{g}L t c_{vw}}{2})}{N\sin(\frac{\gamma \texttt{g}L t c_{vw}}{2}\frac{1}{N-1})} \approx
\frac{\sin(\frac{\gamma \texttt{g}L t c_{vw}}{2})\sin(\frac{\gamma \texttt{g} Lt c_{vw}}{2})}{\frac{\gamma \texttt{g} Lt c_{vw}}{2}} = 
\sinc(\frac{\gamma \texttt{g}L t c_{vw}}{2})\sin(\frac{\gamma \texttt{g} Lt c_{vw}}{2}), \label{eq:imaginary}
\end{eqnarray}
\end{widetext}
Thus, verifying that eq. \eqref{eq:discrete} converges to \eqref{eq:continous2} for large values of $N$.

\begin{figure}%
        \includegraphics[width=0.47\textwidth]{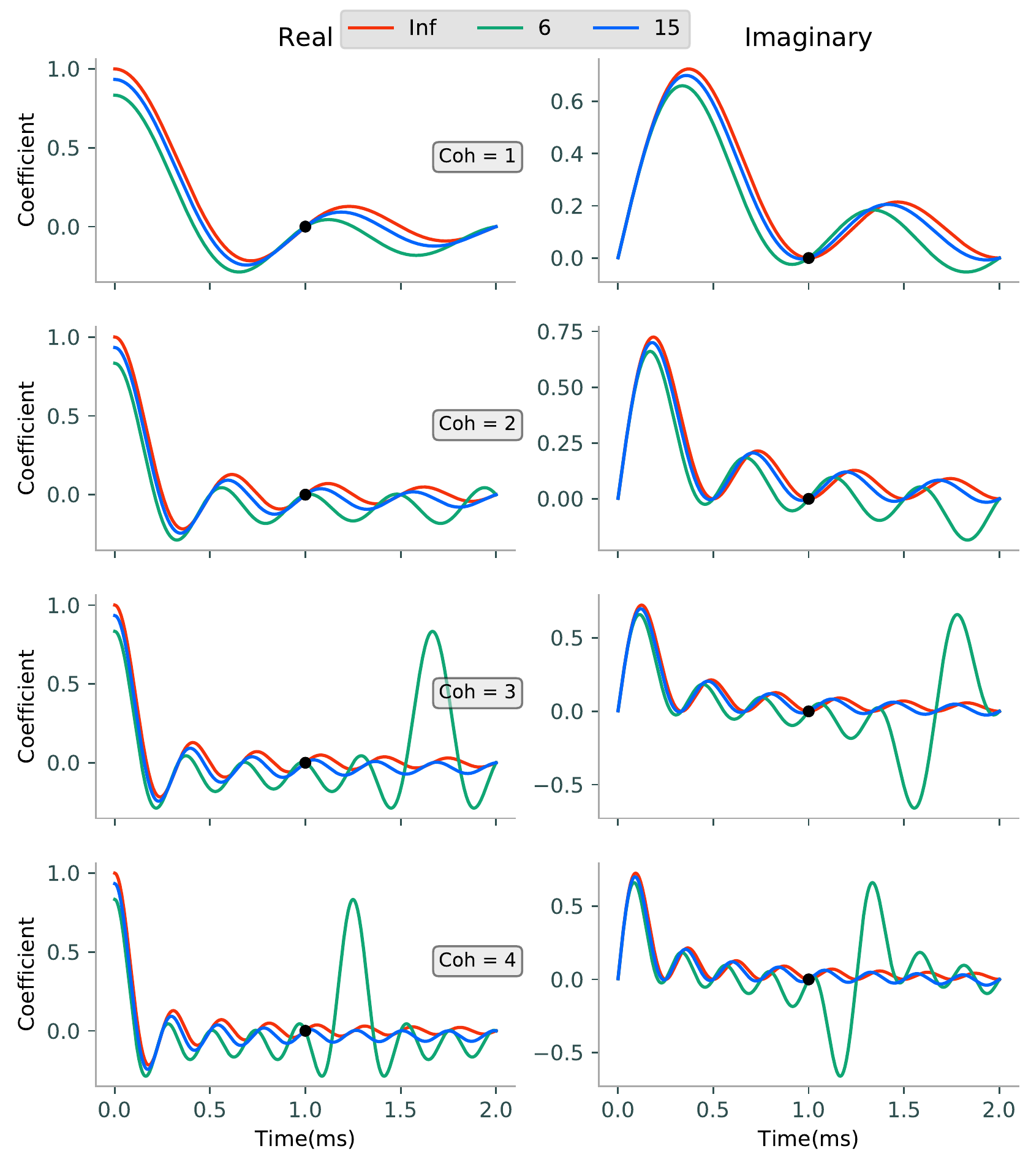}
    \centering
    \caption{Plots of the evolution of the real (left column)  and imaginary (right column) coefficients of coherence order, $1$ to $4$ (top to bottom).
    We compare the continuous case with infinitely many divisions (red) with $N = 6$, and $14$ divisions. It is evident that all the coefficients goes to zero
    at time, $1$ ms, marked by a black dot.}%
    \label{fig:SimCoh}%
\end{figure}

Generally, gradients are used to suppress all the coherences except the zero order, which are unaffected. To choose the minimum number of divisions $N$ for a simulation, we must guarantee that eq. (\ref{eq:discrete}) and eq. (\ref{eq:continous2}) produce close final states.
For $\gamma \texttt{g} L = 2 \pi $ kHz and $\tau = 1$ ms, all the coherences (except zero order) vanishes according to eq. (\ref{eq:continous2}). In this case, it can be seen that for $N>Q+1$, where $Q$ is the number of qubits, all the coherences vanishes from eq. (\ref{eq:discrete}). Since $-Q\leq c_{vw} \leq Q$, for any $N \leq Q+1 $, there will always be one coherence for which denominator of eq. (\ref{eq:discrete}) goes to zero. Thus, for specific configurations, we can simulate the gradient with high precision using a small value for $N$. We plot the evolution of the coefficients of various coherences
with time for aforementioned value of $\gamma \texttt{g} L$ in fig. \ref{fig:SimCoh}. At time $\tau = 1$ ms, we see that the all the coherences are zero irrespective 
of the number of divisions.

\begin{figure}[h]%
    \includegraphics[width=8.5cm]{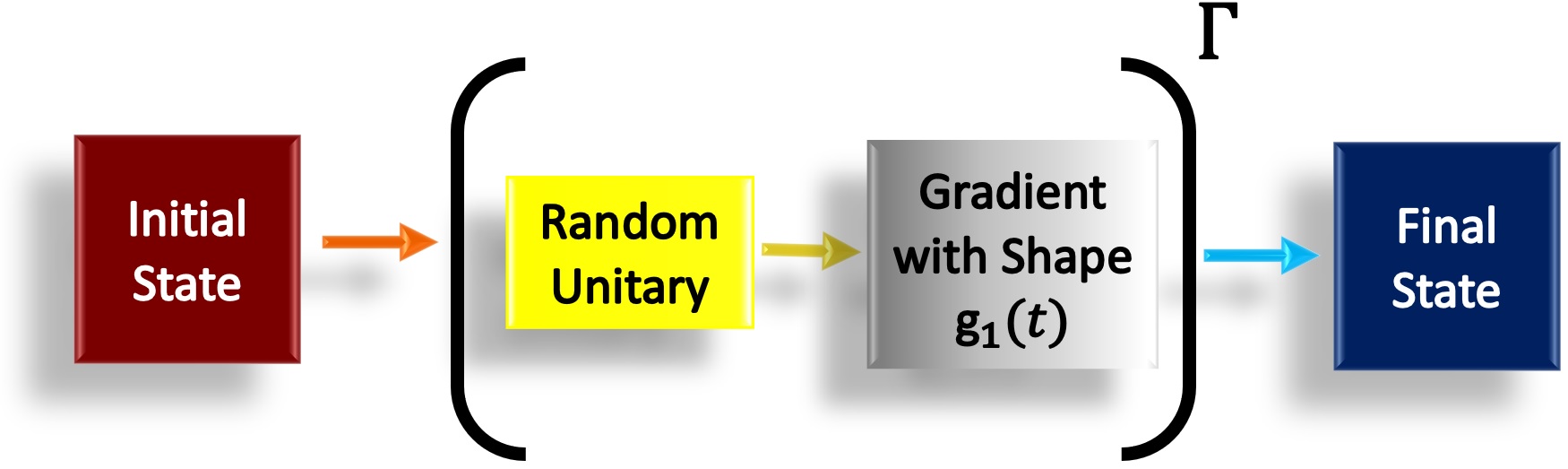}
    \centering
    \caption{Sequence composed of $\Gamma$ repetition of random unitary evolutions and magnetic field gradients with $\tau = 1$ ms. 
    The magnetic field gradient is same for all the repetitions }%
    \label{fig:seqgr2}%
\end{figure}

\begin{figure}[h]%
    \includegraphics[width=8.5cm]{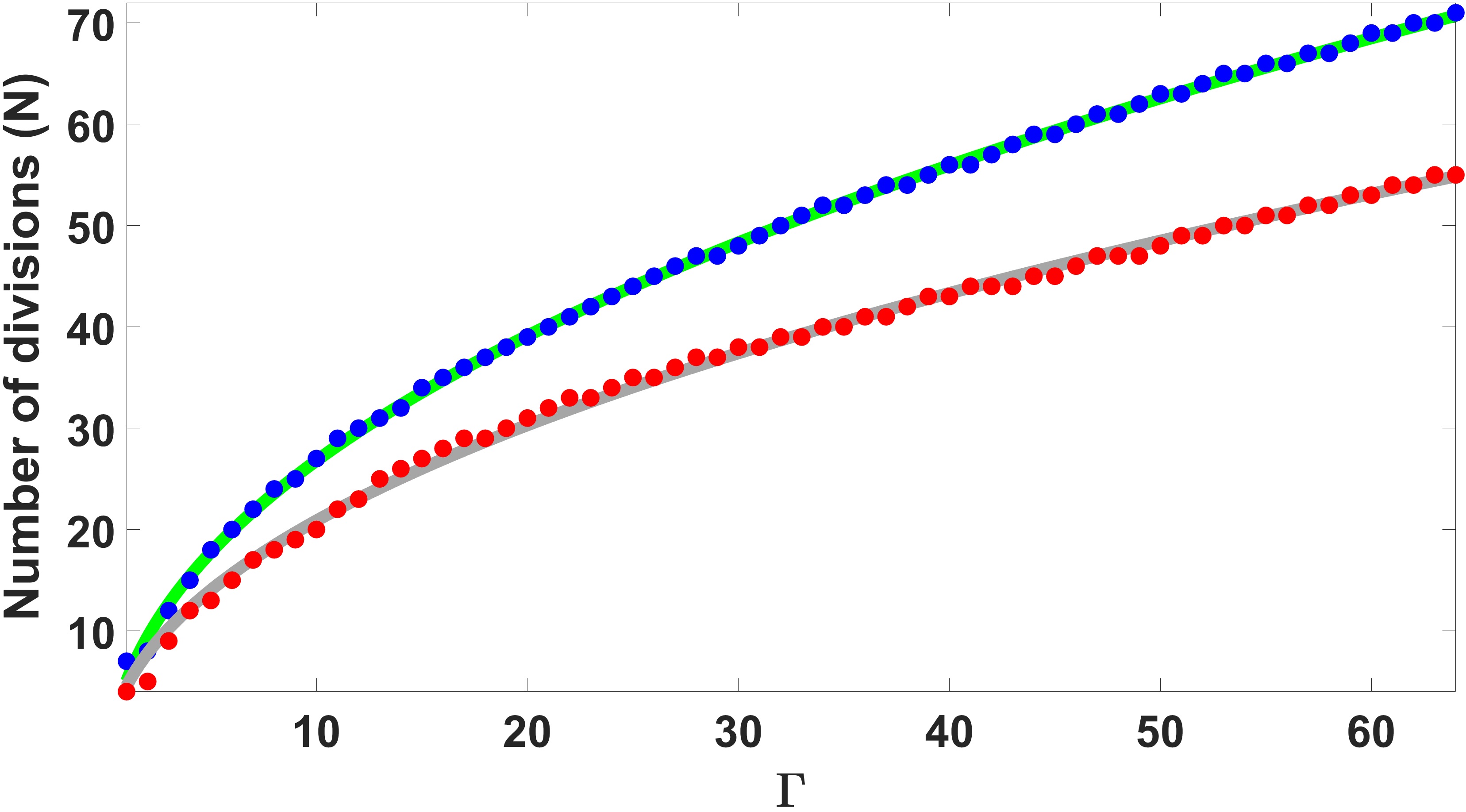}
    \centering
    \caption{Number of divisions needed to simulate the sequence shown in fig. \ref{fig:seqgr2} and obtain a fidelity greater than 0.99999 for different values of $\Gamma$ - The red and blue dots represent the results for the 4 and 7 qubits system, respectively. The lines are fits of the function $N(\Gamma,Q) = a\Gamma^{b} -Q$, where $Q$ is the number of qubits in the system. For the 4 qubtis system, gray line, we obtained that $ a = 8.5 \pm 0.2$ and $ b = 0.464 \pm 0.008$. For the 7 qubtis system, green line, we obtained that $ a = 12.03 \pm 0.04$ and $ b = 0.4486 \pm 0.0008$.}%
    \label{fig:ffid}%
\end{figure}

Now, we verify how $N$ increases, when we simulate a sequence composed of several pulses and gradients. 
For this, we simulated the sequence shown in fig. \ref{fig:seqgr2}, composed of $\Gamma$ random unitary evolutions and magnetic field gradients, fixing $\gamma \tau L  \int_{0.2}^{\tau+0.2}\texttt{g}_{1}(t)dt = 2\pi$ and varying the value of $\delta z$. We used the initial states $\left | 0000 \right  \rangle$ and $\left | 0000000 \right  \rangle$ for the simulation with the 4 and 7 qubits system. The sequence from fig. \ref{fig:seqgr2} was simulated 64 times for each value of $\Gamma$ with a different set of random unitary, and in each simulation the value of $\delta z$ was optimized to obtain a fidelity greater than 0.99999 using the smallest number of molecules. After this optimization, we determined the highest value of $N$ obtained for different values of $\Gamma$. The results for the 4 and 7 qubits systems are presented in fig. \ref{fig:ffid}. We used these results to fit the function $N(\Gamma,Q) = a\Gamma^{b} -Q$. The values of $a$, $b$ and the fitted curves for the 4 and 7 qubits systems are shown in fig. \ref{fig:ffid}. As the values of $a$ and $b$ are small, the value of $N$ will not increase too fast when $\Gamma$ increases. Then, we can still perform a fast simulation with good precision considering an ensemble composed of few molecules, if we choose well the value of $\delta z$. Furthermore, we can use the function $N(\Gamma,Q)$ to have an estimation for the value of $N$ needed to simulate a sequence composed of $\Gamma$ unitary and gradients.

\section{Experiment}\label{sec:experiment}
Here, we report some experimental tests showing that our simulations with a small $N$ agree with the experimental data. The experiments were performed using a Bruker Avance III $700$ MHz NMR spectrometer with the sample containing $^{13}\textrm{C}$-labelled transcrotonic acid dissolved in acetone, the 4 qubits nuclear spins systems from fig. \ref{fig:molecula4q}(a),
at room temperature of 298 K. 

\subsection{Implementing quantum channels}

\begin{figure}[h]%
        \hspace*{-0.4cm}
    \includegraphics[width=6.7cm]{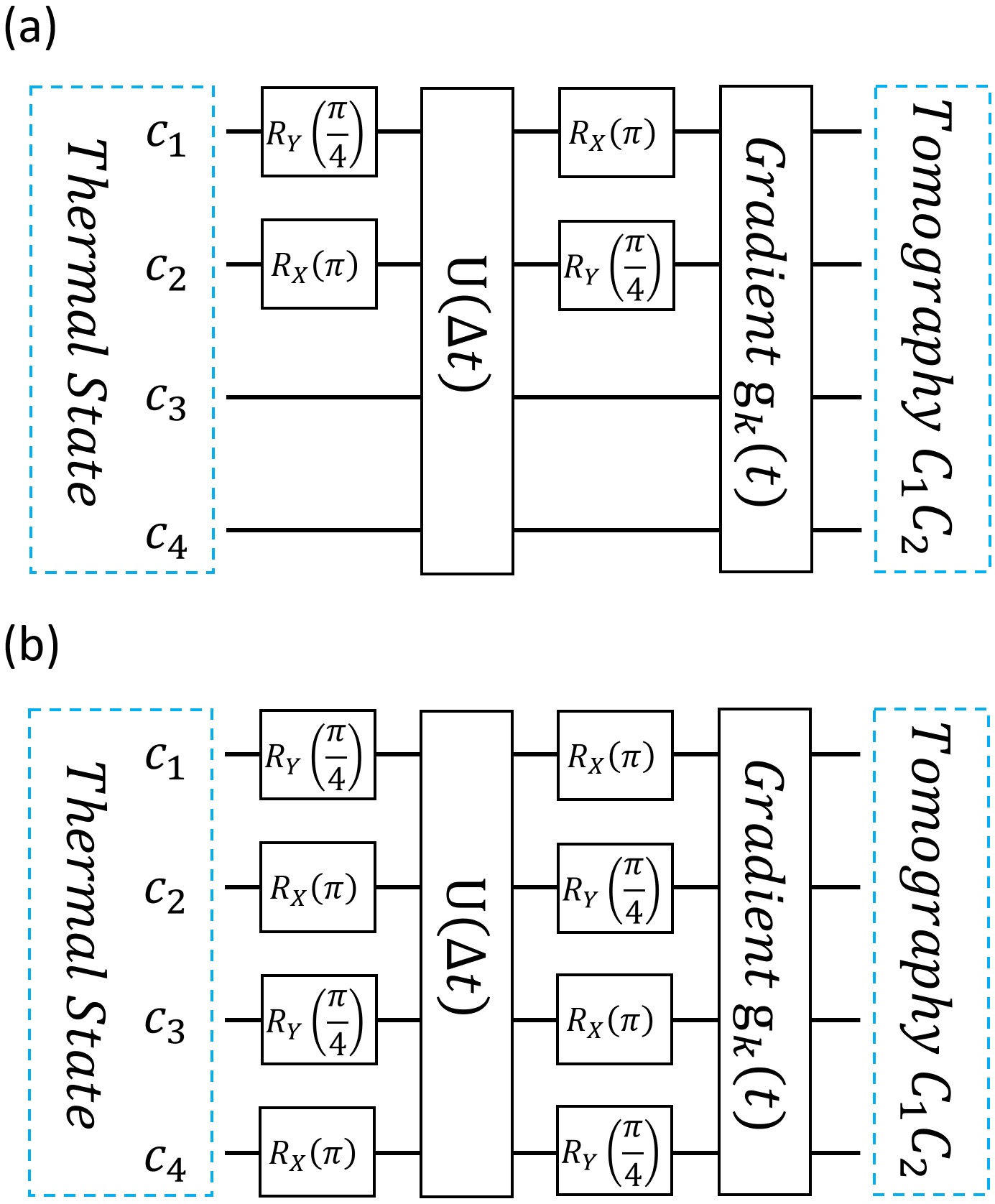}
    \centering
    \caption{Sequences used in the experiments to test if our simulation agree with the experimental data - The two sequence are used to create coherence. Then, the gradient is applied for a time $\tau$ and the state of $C_{1}$ and $C_{2}$ are determined using the quantum state tomography. $U(\Delta t)$ represents a free evolution for a time $\Delta t = 3.459$ ms under the action of $\mathcal{H}_{0}$.}%
    \label{fig:seqtomo}%
\end{figure}

In the first experimental tests, we implemented the two quantum channels shown in fig. \ref{fig:seqtomo}(a-b).
The experiments were performed for two different shapes of gradient, $\texttt{g}_{1}(t)$ or $\texttt{g}_{2}(t)$ from fig. \ref{fig:seqgr}(b), 
with different maximum amplitude and duration of the gradient. 
The pulses used to implement the unitaries were optimized using the technique developed by Peterson \textit{et. al.} \cite{johnp}.
Since in quantum state tomography (QST), the number of measurements increases exponentially with the size of the system,
we performed the QST on the subsystem of two-qubit, $C_{1}$ and $C_{2}$ \cite{livronc,tomografia}. 
By doing so, we reduced the number of measurements for QST and were able to get information about coherence terms of order $0$, $1$, and $2$, and test if our simulations can describe the experiments well.

\begin{figure}[h]%
    \includegraphics[width=8.5cm]{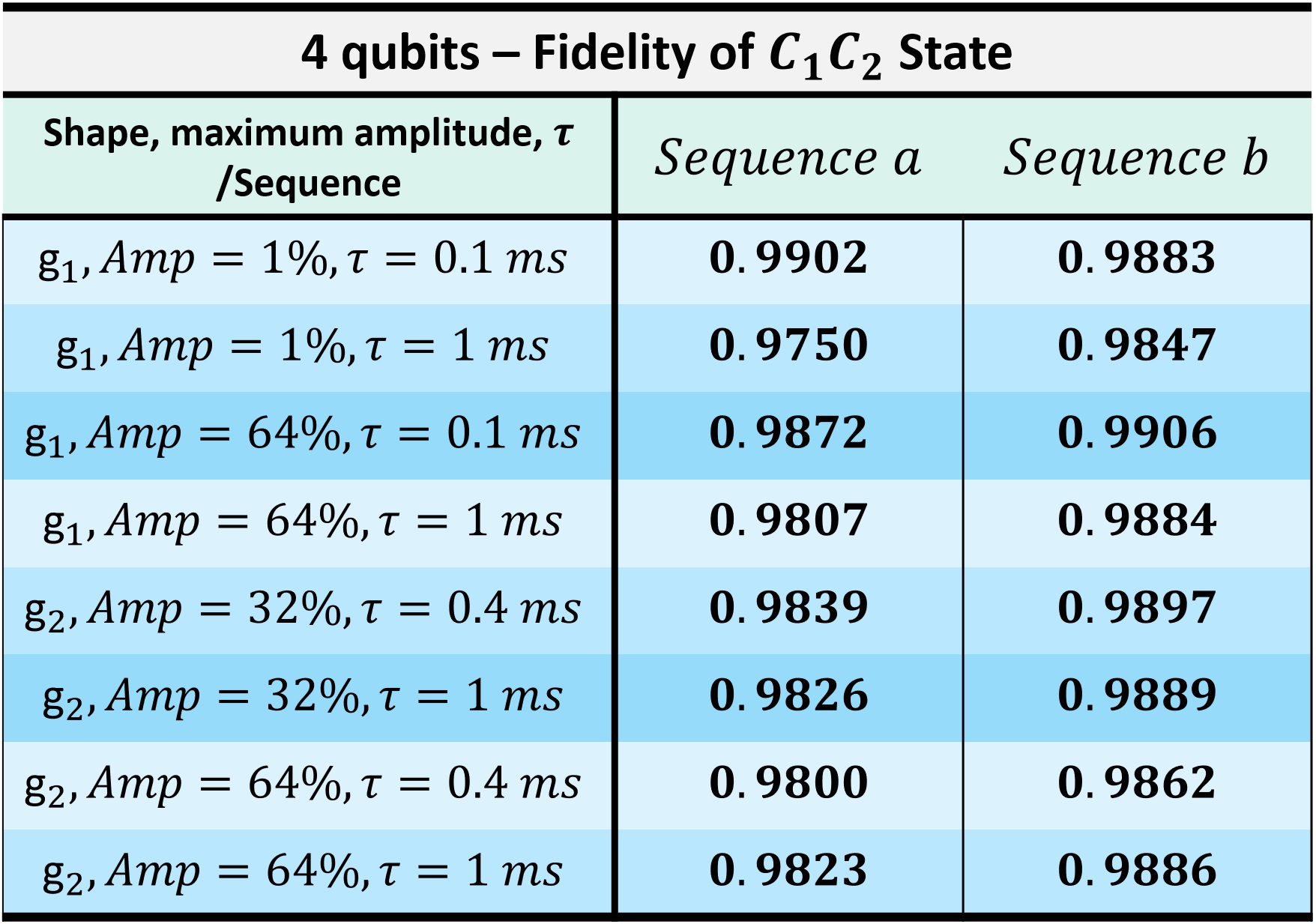}
    \centering
    \caption{ Experimental fidelity -  $Amp = $ max$[\texttt{g}_{k}(t)]$/$G$, where $G$ is the maximum amplitude of the gradient that the experimental equipment can produce. For our set-up, we have $\gamma_{k} L G = 48.6 \pm 0.2$ kHz.}%
    \label{fig:restomo}%
\end{figure}

In fig. \ref{fig:restomo}, we report the fidelity between the experimental state and the simulated (obtained using an ensemble with $N =6$) 
for different types of tests. The fidelity for the simulation with big and small value of $N$ have a value of at least 0.99999, when the value of the space discretization ($\delta z$) is optimized. Our best experimental fidelity is around 0.99. We have slightly worse fidelity experimentally, because in the experiments there are other effects that influence the dynamics of the system, and they are not included in our simulations. The main contributions for these errors are from: the diffusion process, inhomogeneity of the magnetic field
that can cause some extra gradients \cite{livrole}, gradient of temperatures in the sample \cite{atemp}, the optimized pulses, and the field gradients 
not being implemented correctly. Even with these errors, the fidelity obtained is good enough to allow us to use the method presented in this article,
together with an optimization algorithm, to find sequences (composed of pulses, field gradient and free evolutions) to implement a specific non-unitary dynamics that can be use to prepare a specific state or implement a quantum channel \cite{livronc}. 

\subsection{Optimization of a PPS sequence}

One of the applications of the gradient in quantum computation is the preparation of pseudo pure states.
By combining the results of this work with those presented in \cite{johnp} and \cite{oti} with some modifications, 
we were able to obtain sequences to prepare pseudo pure states for 4 and 7 qubit systems. 
These sequences produce pseudo pure state using multiple scan, which corresponds to do multiple experiments (scans) 
each with a similar or different pulse sequence and the result is the average over all the scans.
We set the limit of one magnetic field gradient per scan. Our sequences have a small number of pulses that implement rotations, and produce pseudo pure states with better signal to noise ratio. In order to study the increase in the signal to noise ratio, 
we compared the thermal state spectrum with the pseudo pure states spectrum using the same number of scans.
This implies, when comparing the signal of the pseudo pure state with the thermal state, we are not taking into account
the signal increase resulting from several scans.

In our simulations with the 4 qubits system, we obtained sequences that can double the value of the signal to noise ratio compared to the thermal state. 
In the 7 qubits system, the improvement is, approximately, 3.5 times. By performing simulations with other homonuclear systems with nuclear spins $1/2$,
we note a trend in the maximum increase in the signal to noise ratio: for a system with $Q$ homonuclear spins, 
the signal to noise ratio can be increased, approximately, by a factor of $Q/2$.

\begin{figure}[h]%
    \includegraphics[width=8.5cm]{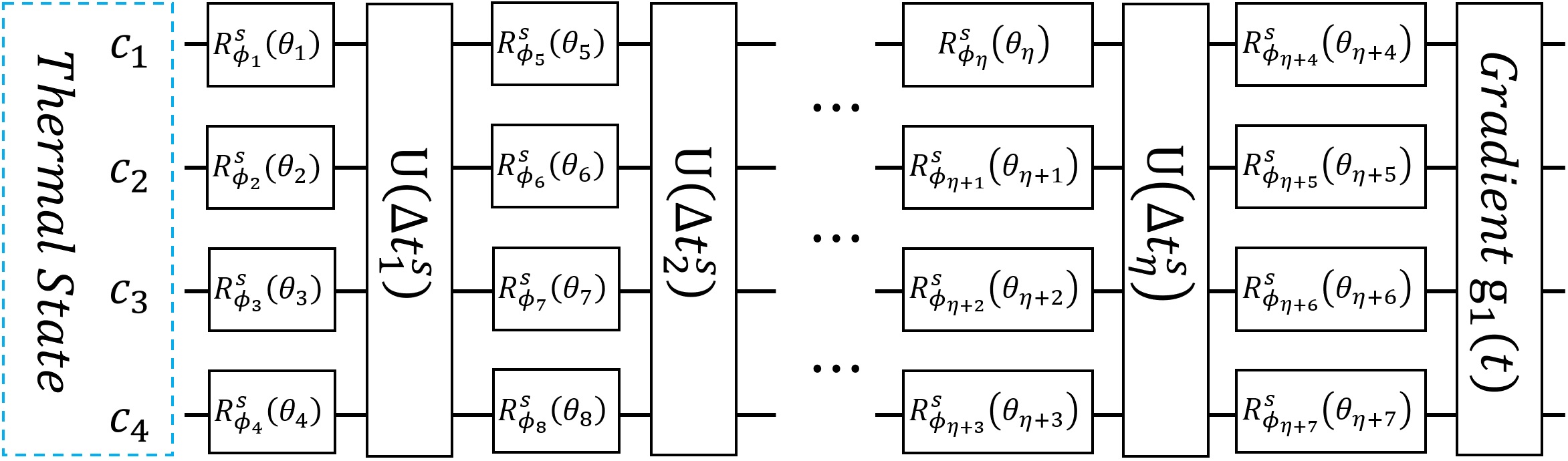}
    \centering
    \caption{Sequence of rotations, free evolutions and gradient used in the optimization to prepare a pseudo pure state - The index $s$ represents the set of angles and times used in the scan $s$, $U(\Delta t)$ represents a free evolution for a time $\Delta t$ under the action of $\mathcal{H}_{0}$. The $R_{\phi}^{s}(\theta)$ is a rotation of an angle $\theta$, around the axis $\zeta = cos(\phi)\hat{x} + sin(\phi)\hat{y}$.  }%
    \label{fig:otimi}%
\end{figure}

In our algorithm, the time of the free evolutions, the angles and phases of the rotations of the circuit presented in fig. \ref{fig:otimi} 
are optimized to minimize the value of the function:
\begin{equation}\label{eq:fide}
\begin{split}
\ \mathcal{F} =  [1-\texttt{Fidelity}(\rho_{f},\rho_{pps})](1-\epsilon) + \epsilon \left \| Q/2 -M \right \|,
\end{split}
\end{equation}
where $\rho_{pps}$ is the theoretical pseudo pure state, $\rho_{f}$ is the final state obtained after the simulation of the circuit 
presented in fig. \ref{fig:otimi}, $M$ is the element of $\rho_{f}$ that has the highest absolute value and $\epsilon$ is a number, 
in the interval $[0,0.5]$, that can be used to prioritize in the optimization the fidelity or the improvement in the signal to noise ratio 
of the pseudo pure state. In our optimization, we used the shape $\texttt{g}_{1}(t)$ for the gradient, with $\tau = 1$ ms.

\begin{figure}[h]%
    \includegraphics[width=7.5cm]{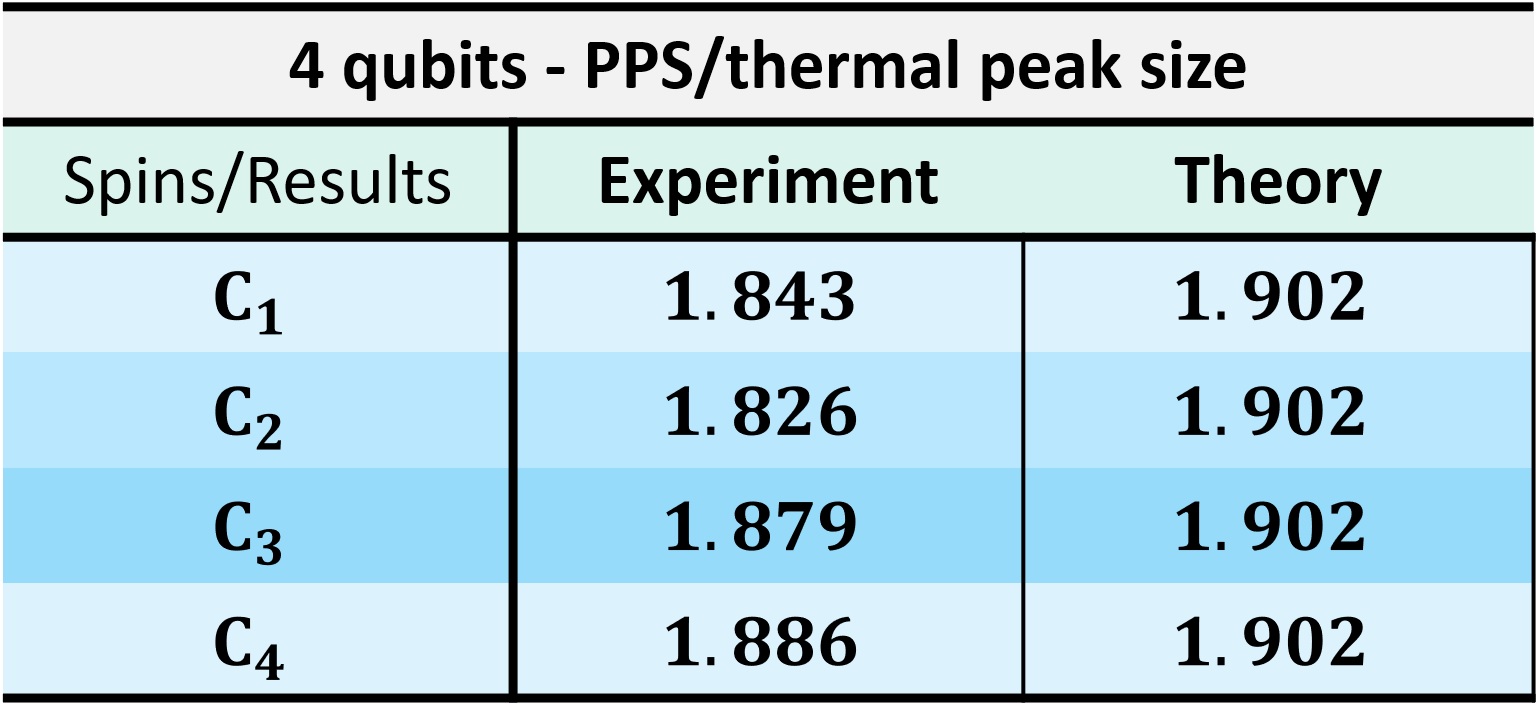}
    \centering
    \caption{ Improvement in the signal to noise ratio using our optimal sequence to prepare the pseudo pure state with the 4 qubits system. }%
    \label{fig:ppsinte}%
\end{figure}

 As a final test for our algorithm, we optimized the angles and times of the sequence from fig. \ref{fig:otimi} to prepare a pseudo pure state for the 4 qubits system. Instead of search for a sequence that give an improvement of 2 in the signal to noise ratio, we prioritized the length of the sequence and the fidelity. To make the sequence short, we used 2 scan, and in each one we have a different set of angles and times of free evolutions. Then, in theory, the improvement in the signal to noise ratio is 1.902, the theoretical fidelity is higher than 0.9999, the duration of the sequence is, approximately, $27$ ms and for each scan we fix $\eta = 5$ in the circuit shown in fig. \ref{fig:otimi}. The improvement in the signal to noise ratio measured experimentally is presented in fig. \ref{fig:ppsinte}. The fidelity of the pseudo pure state prepared experimentally is 0.998.

\section{Conclusions}
Pulse field gradients (PFG) are important for different areas of science. In physics, they are essential to prepare certain states, perform measurements, measure the diffusion coefficient of a sample, and implement non-unitary dynamics (quantum channels). We studied how to efficiently simulate PFGs using discretization in time and space. Utilizing the recent techniques developed by Bhole and Jones \cite{apro}, we provide
a guideline for the size of time discretization, depending upon the fidelity required in the experiment. We show how to efficiently
discretize the space in a very small number of divisions, and still be able to simulate the PFGs with high precision. We show that for a system of $Q$ qubits, when applying a single gradient, the minimum number of divisions needed in space is $Q+2$. For sequences composed of multiple gradients and unitary evolutions, we shown with our simulations that number of slices 
vary as $a\Gamma^b - Q$, where $\Gamma$ is the number of repetitions of gradients and unitary evolutions implemented, $a$ and $b$ are small numbers and depend upon the system being studied. As the number of slices is small, and do not increases too fast when $\Gamma$ increases, we can simulate the sequences quickly and with high precision when the value of the space discretization is optimized. 

We perform two types of experiments which utilize the above developed techniques. Firstly, we implemented two quantum channel and determine part of the state of the system to compare with the theoretical prediction. For all states determined experimentally, the fidelity was higher than 0.98. In the second experiment, we show that the method presented here to simulate the gradient, can be used together with an optimization algorithm to find the optimum sequence to prepare the pseudo pure state. We describe how to perform the optimization to obtain a sequence to prepare a pseudo pure state with better signal to noise ratio, compared to any other procedure we are aware of. We were able to see a trend in our simulations showing that $Q/2$ is the maximum improvement of the signal to noise ratio, for a homonuclear system of $Q$ qubits. At the end, we implemented a optimized sequence to prepare the pseudo pure state in four qubits systems, and measure, experimentally, a improvement higher than 1.8 in the signal to noise ratio. The fidelity of the experimental state is higher than 0.99. Thus, in addition to demonstrate that a fast simulation of the dynamics of a NMR sample state under the influence of field gradient is possible, we compared our fast simulation results with the experimental data, and we show how a fast simulation can be applied to optimize sequences. 
The results of this study was already used to design a quantum Szilard engine, which uses information about the state of a system to fully convert heat into work \cite{szi}. We believe that our study can help in  designing optimum NMR sequences, composed of PFGs, employed in different areas of science.
  
\begin{acknowledgments}
 We acknowledge financial support from Ministery of Innovation,
Science and Economic Development (Canada), the Government of Ontario, CIFAR, Mike and Ophelia Lazaridis.
\end{acknowledgments}

\bibliography{RefGradients}

\appendix
\section{A1}\label{ap:a1}
In case of homonuclear sample and time independent gradient:
\begin{eqnarray}
\mathcal{H}_{\texttt{g}} &=& \frac{ \gamma \texttt{g} z}{2} \sum_{k}\sigma_{k}^{z}, \ \  \texttt{and} \\
U_{\texttt{g}}(z) &=& \expc(-i \mathcal{H}_{\texttt{g}} t/\hbar).
\end{eqnarray}

Evaluation evolution under the gradient:
\begin{eqnarray}
        \sigma_{k}^{z}\ket{b_v} &=& (-1)^{b_v^k}\ket{b_v} \nonumber \\
\Rightarrow \sum_k \sigma_{k}^{z}\ket{b_v} &=& \sum_k (-1)^{b_v^k} \ket{b_v} \nonumber \\
\Rightarrow \expc\left(-i\frac{ \gamma \texttt{g} z t}{2} \sum_{k}\sigma_{k}^{z}\right)\ket{b_v} &=& \expc\left(-i\frac{ \gamma \texttt{g} z t}{2} \sum_{k} (-1)^{b_v^k}\right)\ket{b_v} \nonumber \\ 
        &=& U_{\texttt{g}}(z) \ket{b_v}. \label{eqA:gradientevo}
\end{eqnarray}
Thus, evolution of a density matrix term under the gradient at position $z$ and time $t$ is:
\begin{eqnarray}
        U_{\texttt{g}}(t,z)\outpr{b_v}{b_w}&& U_{\texttt{g}}(t,z)^{\dagger} \nonumber \\
        &&= \expc\left(-i\frac{ \gamma \texttt{g} z t}{2} \sum_{k}[(-1)^{b_w^k} - (-1)^{b_v^k}]\right) \outpr{b_v}{b_w} \nonumber \\
        &&= \expc\left(-i\gamma \texttt{g} z t c_{vw}\right) \outpr{b_v}{b_w}. \label{eq:grevo}
\end{eqnarray}
In the last line we have used eq. (\ref{eq:coh}).
To solve for $\rho_{S}(t)$, we consider the sample of length $L$ and $z \in [0,L]$, and solve for the individual elements of eq. (\ref{eq:estadot1}):
\begin{widetext}
\begin{eqnarray}
&& \int_{0}^{L}\expc\left(-i\gamma \texttt{g} z t c_{vw}\right) dz \\
&=& \frac{\expc\left(-i\gamma \texttt{g} z t c_{vw}\right) \vert^{L}_{0}}{-i\gamma \texttt{g}t c_{vw}} \\
&=& \frac{\expc\left(-i\gamma \texttt{g} L t c_{vw}\right) -1 }{-i\gamma \texttt{g}t c_{vw}} \\
&=& \frac{\expc\left(-i\gamma \texttt{g} L t c_{vw}/2\right)\Big(\expc\left(-i\gamma \texttt{g} L t c_{vw}/2\right) -\expc\left(i\gamma \texttt{g} L t c_{vw}/2\right)\Big) }{-i\gamma \texttt{g}t c_{vw}} \\
&=& \frac{\expc\left(-i\gamma \texttt{g} L t c_{vw}/2\right)\Big(2 \sin(\gamma \texttt{g} L t c_{vw}/2)\Big) }{\gamma \texttt{g}t c_{vw}} \\
&=& \frac{2\cos(\gamma \texttt{g} L t c_{vw}/2)\sin(\gamma \texttt{g} L t c_{vw}/2) - i 2 \sin^2(\gamma \texttt{g} L t c_{vw}/2)}{\gamma \texttt{g}t c_{vw}} \\
&=& \frac{\sin(\gamma \texttt{g} L t c_{vw}) - i 2 \sin^2(\gamma \texttt{g} L t c_{vw}/2)}{\gamma \texttt{g}t c_{vw}}.
\end{eqnarray}
Also, $\int dz = L$, thus:
\begin{eqnarray}
\rho_{S}(t) &=& \sum_{v,w\in [1,2^{Q}]} \frac{a_{vw} A_{vw}\Big(\frac{\sin(\gamma \texttt{g} L t c_{vw}) - i 2 \sin^2(\gamma \texttt{g} L t c_{vw}/2)}{\gamma \texttt{g}t c_{vw}}\Big)\outpr{b_v}{b_w}}{L} \\
&=& \sum_{v,w\in [1,2^{Q}]} a_{vw} A_{vw}\Big(\frac{\sin(\gamma \texttt{g} L t c_{vw}) - i 2 \sin^2(\gamma \texttt{g} L t c_{vw}/2)}{\gamma \texttt{g} L t c_{vw}}\Big)\outpr{b_v}{b_w} \\
&=& \sum_{v,w\in [1,2^{Q}]} a_{vw} A_{vw}\Big(\frac{\sin(\gamma \texttt{g} L t c_{vw})}{\gamma \texttt{g} L t c_{vw}} - i \frac{2 \sin^2(\gamma \texttt{g} L t c_{vw}/2)}{\gamma \texttt{g} L t c_{vw}}\Big)\outpr{b_v}{b_w} \\
&=& \sum_{v,w\in [1,2^{Q}]} a_{vw} A_{vw}\Big(\sinc(\gamma \texttt{g} L t c_{vw}) - i \sinc(\gamma \texttt{g} L t c_{vw}/2)\sin(\gamma \texttt{g} L t c_{vw}/2) \Big)\outpr{b_v}{b_w}. 
\end{eqnarray}

We prove an important identity before proceeding:

If $p_m = p_0 + (m-1)d$, then
\begin{eqnarray} \label{eq:magic}
\sum_{m=1}^{M} \expc(ip_mx) = i\left(\frac{\expc(i(p_0-\frac{d}{2})x)-\expc(i(Md+p_0-\frac{d}{2})x)}{2\sin(\frac{dx}{2})}\right). 
\end{eqnarray}

\textit{Proof:} 
\begin{eqnarray}
\sum_{m=1}^{M} \expc(ip_mx) &=& \sum_{m=1}^{M} \expc(i(p_0 + (m-1)d)x) \nonumber \\
&=& \expc(ip_0x) \sum_{m=1}^{M} \expc(i(m-1)dx) \label{eq:gp1}\\
&=& \expc(ip_0x) \left(\frac{1-\expc(iMdx)}{1-\expc(idx)}\right) \label{eq:gp2} \\
&=& \expc(ip_0x) \left(\frac{1-\expc(iMdx)}{1-\expc(idx)}\right)\frac{\expc(-i\frac{d}{2}x)}{\expc(-i\frac{d}{2}x)} \nonumber \\
&=& \expc(i(p_0-\frac{d}{2})dx) \left(\frac{1-\expc(iMdx)}{\expc(-i\frac{d}{2}x)-\expc(i\frac{d}{2}x)}\right) \nonumber \\
&=& i\left(\frac{\expc(i(p_0-\frac{d}{2})x)-\expc(i(Md+p_0-\frac{d}{2})x)}{2\sin(\frac{dx}{2})}\right). \nonumber 
\end{eqnarray}
Going from eq.(\ref{eq:gp1}) to eq.(\ref{eq:gp2}), we have made use of the fact that it is a summation of a geometric series.

Similar to continuous case, we consider a sample of length $L$, $z_m = (m-1)\frac{L}{N-1}$ and solve for individual elements of
eq.(\ref{eq:estadot}),
\begin{eqnarray}
&& \sum_{m=1}^{N} a_{vw} A_{vw}\expc\left(-i\gamma \texttt{g} z_m t c_{vw}\right) \outpr{b_v}{b_w}dz \nonumber \\
&=& a_{vw} A_{vw}\outpr{b_v}{b_w}\sum_{m=1}^{N} \expc\left(-i\gamma \texttt{g} z_m t c_{vw}\right), \nonumber \\
&&\text{making use of eq.(\ref{eq:magic}) with }p_m = z_m, p_0 = 0, d = \frac{L}{N-1} \text{ and } x = -\gamma \texttt{g} t c_{vw}, \nonumber \\
&=& a_{vw} A_{vw}\outpr{b_v}{b_w}i\left(\frac{\expc(i(-\frac{L}{2(N-1)})x)-\expc(i(N\frac{L}{N-1}-\frac{L}{2(N-1)})x)}{2\sin(\frac{L}{2(N-1)}x)}\right),  \\
&& \text{ let } \alpha = -\frac{xL}{2(N-1)}  \ \text{ and }\  \beta = \frac{xL(2N-1)}{2(N-1)} \\
&=& a_{vw} A_{vw}\outpr{b_v}{b_w}i\left(\frac{\expc(i\alpha)-\expc(i\beta)}{2\sin(\frac{xL}{2(N-1)})}\right) \\
&=& a_{vw} A_{vw}\outpr{b_v}{b_w}i\left(\frac{\cos(\alpha)-\cos(\beta) + i (\sin(\alpha)-\sin(\beta))}{2\sin(\frac{xL}{2(N-1)})}\right) \\ 
&=& a_{vw} A_{vw}\outpr{b_v}{b_w}i\left(\frac{-2\sin(\frac{\alpha+\beta}{2})\sin(\frac{\alpha-\beta}{2}) + i 2\cos(\frac{\alpha+\beta}{2})\sin(\frac{\alpha-\beta}{2})}{2\sin(\frac{xL}{2(N-1)})}\right). \label{eq:continue}
\end{eqnarray}

Let's simplify a bit, 
\begin{eqnarray}
&& \frac{\alpha+\beta}{2} = \frac{1}{2}\Big(-\frac{xL}{2(N-1)} + \frac{xL(2N-1)}{2(N-1)}\Big) = \frac{xL}{2}\frac{N}{N-1}, \\
&& \frac{\alpha-\beta}{2} = \frac{1}{2}\Big(-\frac{xL}{2(N-1)} - \frac{xL(2N-1)}{2(N-1)}\Big) = \frac{-xL}{2}\frac{N-1}{N-1} = \frac{-xL}{2}. 
\end{eqnarray}

Continuing eq. (\ref{eq:continue}),
\begin{eqnarray}
&=& a_{vw} A_{vw}\outpr{b_v}{b_w}i\left(\frac{-2\sin(\frac{xL}{2}\frac{N}{N-1})\sin(\frac{-xL}{2}) + i2\cos(\frac{xL}{2}\frac{N}{N-1})\sin(\frac{-xL}{2})}{2\sin(\frac{xL}{2}\frac{1}{N-1})}\right) \\
&=& a_{vw} A_{vw}\outpr{b_v}{b_w}\left(\frac{\cos(\frac{xL}{2}\frac{N}{N-1})\sin(\frac{xL}{2})}{\sin(\frac{xL}{2}\frac{1}{N-1})} + i
\frac{\sin(\frac{xL}{2}\frac{N}{N-1})\sin(\frac{xL}{2})}{\sin(\frac{xL}{2}\frac{1}{N-1})}
\right)  \\
&=& a_{vw} A_{vw}\outpr{b_v}{b_w}\left(\frac{\cos(\frac{\gamma \texttt{g}L t c_{vw}}{2}\frac{N}{N-1})\sin(\frac{\gamma \texttt{g} Lt c_{vw}}{2})}{\sin(\frac{\gamma \texttt{g} Lt c_{vw}}{2}\frac{1}{N-1})} + i
\frac{\sin(\frac{\gamma \texttt{g} Lt c_{vw}}{2}\frac{N}{N-1})\sin(\frac{\gamma \texttt{g}L t c_{vw}}{2})}{\sin(\frac{\gamma \texttt{g}L t c_{vw}}{2}\frac{1}{N-1})}
\right).
\end{eqnarray}

Therefore,
\begin{eqnarray}
\rho_{S}(t) &=& \sum_{v,w\in [1,2^{Q}]}a_{vw} A_{vw}\outpr{b_v}{b_w}\left(\frac{\cos(\frac{\gamma \texttt{g}L t c_{vw}}{2}\frac{N}{N-1})\sin(\frac{\gamma \texttt{g} Lt c_{vw}}{2})}{\sin(\frac{\gamma \texttt{g} Lt c_{vw}}{2}\frac{1}{N-1})} + i
\frac{\sin(\frac{\gamma \texttt{g} Lt c_{vw}}{2}\frac{N}{N-1})\sin(\frac{\gamma \texttt{g}L t c_{vw}}{2})}{\sin(\frac{\gamma \texttt{g}L t c_{vw}}{2}\frac{1}{N-1})}
\right)/N \nonumber \\
&=& \sum_{v,w\in [1,2^{Q}]}a_{vw} A_{vw}\outpr{b_v}{b_w}\left(\frac{\cos(\frac{\gamma \texttt{g}L t c_{vw}}{2}\frac{N}{N-1})\sin(\frac{\gamma \texttt{g} Lt c_{vw}}{2})}{N\sin(\frac{\gamma \texttt{g} Lt c_{vw}}{2}\frac{1}{N-1})} + i
\frac{\sin(\frac{\gamma \texttt{g} Lt c_{vw}}{2}\frac{N}{N-1})\sin(\frac{\gamma \texttt{g}L t c_{vw}}{2})}{N\sin(\frac{\gamma \texttt{g}L t c_{vw}}{2}\frac{1}{N-1})}
\right). \nonumber
\end{eqnarray}

\end{widetext}

\end{document}